\documentclass[12pt,a4paper]{book}
\usepackage{npcs}
\usepackage{graphics} 
\usepackage{epsfig}

\ntitle{Non-zero Quark Modes Contribution to the
QCD-Instanton-Induced  Deep Inelastic Scattering
}

\nauthor{V.I.Kashkan$^1$, V.I.Kuvshinov$^2$,
R.G.Shulyakovsky$^3$}

\naddress{B.I. Stepanov Institute of Physics,\\
 National Academy of Sciences of Belarus \\
\it F. Skorina avenue 68,\\
 220072 Minsk, Belarus \\
{\small \rm
   E-mails:     $^1$kashkan@dragon.bas-net.by \\
   $^2$kuvshino@dragon.bas-net.by \\
   $^3$shul@dragon.bas-net.by \\
}}

\ndata{
 }

\nabstract{Instanton-induced deep-inelastic processes are
considered. Unlike previous results we calculate nonzero quark
modes contribution. }

\nkeywords{ instantons, non-zero quark modes, deep inelastic
scattering }

\nPACSnumbers{ 11.15.Tk }
\begin{document}
\DeclareGraphicsExtensions{.jpg,.pdf,.mps,.png} 
\firstpage{2}  \nyear{2001}
\def\nfpage{\thepage}
\thispagestyle{myheadings} \npcstitle
\begin{sloppypar} As was shown~\cite{Bal}, QCD-instantons can
appear at the intermediate stage of deep inelastic scattering
(DIS) and be, in principle, observed at HERA
(DESY)~\cite{RSh,Acta}. Specifically, instantons can be produced
in quark-gluon subprocess (Fig.1).

\hspace*{3cm}
\begin{minipage}{3.5in}\begin{center}
\epsfxsize=2.5in \epsfysize=2.5in \epsfbox{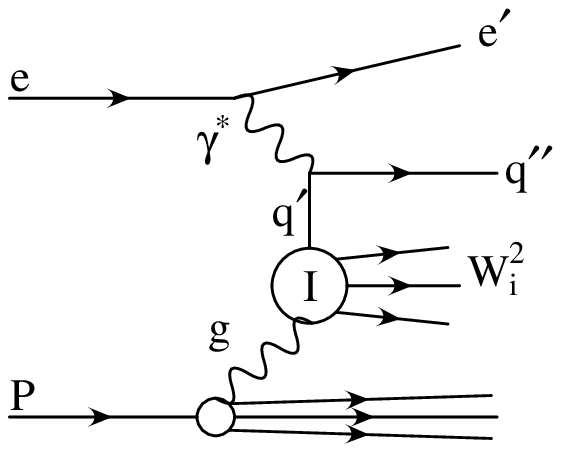}\end{center}
\end{minipage}
\vspace*{-12.6cm}

\vspace*{13cm}
\begin{minipage}{16.5cm}
\begin{center}
Fig.1. Instanton induced DIS.\end{center}
\end{minipage}
 \vspace*{0.1cm}

Usually it is supposed that only minimal number of quarks is
produced after "decay" of instanton. Number of final gluons is
arbitrary:
\begin{equation}\label{process}
q+g\to (2n_f-1)q+n_gg.
\end{equation}
Here we consider instanton-induced processes with arbitrary number
of quarks. Contribution of these processes is determined by
nonzero fermion propagator~\cite{Shuryak}
$$
S^{nz}(x,y)=
\frac{1}{\sqrt{1+\rho^2/x^2}}\frac{1}{\sqrt{1+\rho^2/y^2}}\Biggl[\frac{(x-y)_{\mu}\sigma_{\mu}}{2\pi^2(x-y)^4}
\biggl(1+\rho^2\frac{x_{\nu}\sigma_{\nu}y_{\kappa}\bar{\sigma}_{\kappa}}{x^2y^2}
\biggr)-
$$
\begin{equation}\label{nonzero}-\frac{\rho^2}{4\pi^2(x-y)^2x^2y^2}
\biggl(\bar{\sigma}_{\mu}\frac{x_{\nu}\sigma_{\nu}\bar{\sigma}_{\mu}(x-y)_{\lambda}\sigma_{\lambda}y_{\omega}
\bar{\sigma}_{\omega}}{\rho^2+x^2}+\sigma_{\mu}
\frac{x_{\nu}\sigma_{\nu}(x-y)_{\lambda}\bar{\sigma}_{\lambda}\sigma_{\mu}y_{\omega}
\bar{\sigma}_{\omega}}{\rho^2+y^2}\biggr)\Biggr],
\end{equation}
where $\rho$ is instanton size, $\sigma_{\mu}=(-i\sigma_a,I),\,
\bar{\sigma}_{\mu}=(i\sigma_a,I)$.

After calculation\footnote{for detailed explanation of calculation
of the distribution of multiplicity on quarks produced after
instanton "decay" see~\cite{me}} we obtain Poisson distribution on
number of quark pairs (for every light quarks), which are produced
in the instanton processes:
\begin{equation}\label{Dist}
P_n=[e^{\xi^2}-1]^{-1}\frac{\xi^{2n}}{n!},\quad
\xi\approx\frac{1-x'}{x'},
\end{equation}
where Bjorken variable of instanton subprocess $x'>0.5$. Average
number of quark pairs for small $\xi$ reads
\begin{equation}\label{Dist}
<n>\approx 3(1+\xi^2).
\end{equation}
Contribution of non-zero modes can be important for the
experimental search of QCD-instantons. Monte-Carlo simulation is
in progress.

\end{sloppypar}
\end{document}